\documentclass[sigconf, screen]{acmart}

\AtBeginDocument{%
  }

\setcopyright{acmlicensed}
\acmPrice{15.00}
\acmDOI{10.1145/3611643.3613887}
\acmYear{2023}
\copyrightyear{2023}
\acmSubmissionID{fse23industry-p73-p}
\acmISBN{979-8-4007-0327-0/23/12}
\acmConference[ESEC/FSE '23]{Proceedings of the 31st ACM Joint European Software Engineering Conference and Symposium on the Foundations of Software Engineering}{December 3--9, 2023}{San Francisco, CA, USA}
\acmBooktitle{Proceedings of the 31st ACM Joint European Software Engineering Conference and Symposium on the Foundations of Software Engineering (ESEC/FSE '23), December 3--9, 2023, San Francisco, CA, USA}
\received{2023-05-18}
\received[accepted]{2023-07-31}

\usepackage{multirow}
\usepackage{microtype}

\begin{document}

\title{Issue Report Validation in an Industrial Context}

\author{Ethem Utku Aktas}
\affiliation{%
  \institution{Softtech Inc., Research and Development Center}
  \city{Istanbul}
  \country{Turkey}
}
\email{utku.aktas@softtech.com.tr}

\author{Ebru Cakmak}
\authornote{Contributed to the study while working at Softtech.}
\affiliation{%
  \institution{Microsoft EMEA}
  \city{Istanbul}
  \country{Turkey}}
\email{ebrucakmak@microsoft.com}

\author{Mete Cihad Inan}
\affiliation{%
  \institution{Softtech Inc., Research and Development Center}
  \city{Istanbul}
  \country{Turkey}
}
\email{cihad.inan@softtech.com.tr}

\author{Cemal Yilmaz}	
\affiliation{%
 \institution{Faculty of Engineering and Natural Sciences, Sabanci University}
 \city{Istanbul}
 \country{Turkey}
}
\email{cyilmaz@sabanciuniv.edu}

\renewcommand{\shortauthors}{Ethem Utku Aktas, Ebru Cakmak, Mete Cihad Inan, Cemal Yilmaz}

\begin{abstract}
Effective issue triaging is crucial for software development teams to improve software quality, and thus customer satisfaction. Validating issue reports manually can be time-consuming, hindering the overall efficiency of the triaging process. This paper presents an approach on automating the validation of issue reports to accelerate the issue triaging process in an industrial set-up. We work on $1,200$ randomly selected issue reports in banking domain, written in Turkish, an agglutinative language, meaning that new words can be formed with linear concatenation of suffixes to express entire sentences. We manually label these reports for validity, and extract the relevant patterns indicating that they are invalid. Since the issue reports we work on are written in an agglutinative language, we use morphological analysis to extract the features. Using the proposed feature extractors, we utilize a machine learning based approach to predict the issue reports' validity, performing a $0.77$ F1-score. 
\end{abstract}




\maketitle

\section{Introduction}
\label{sec:intro}

Being a subsidiary of IsBank\footnote{https://www.isbank.com.tr}, Turkey's largest private bank, Softtech\footnote{https://softtech.com.tr} received more than a hundred thousand issue reports from IsBank in 2022, relating to software problems faced by bank clerks or customers in various channels. Softtech serves IsBank with over seven million digital customers who conduct millions of transactions daily through ATM and POS machines, web applications, mobile devices and bank branches. A team of $50$ people at IsBank (IT Help Desk), and $30$ people at Softtech (Application Support Team) work in resolving these reports. To automate and accelarate the issue assignment process at Softtech, we developed and deployed a tool called \textit{IssueTAG} \citep{aktas2020automated, aktas2022using} in February $2018$. 

After assignment, the issue reports are either analyzed and solved by the application support team or developers, or returned to the reporters because they are feature requests, operational requests, questions, etc. The issue triaging process at Softtech could further be accelerated by detecting reports that are not actual issue reports, as $25.8\%$ of the reports resolved by Softtech in $2022$ were returned by the developers or the application support team to the reporters, having an average resolution time of $3.46$ days (Section \ref{sec:motivation}). Note that the average resolution time for closing a report which is not returned is $3.91$ days in the same year, which highlights the need for automated validation of the incoming issue reports for reducing the issue triaging costs at Softtech.


The issue reports that are not related to corrective maintenance of the software are called as $non-issue$ reports in this paper. Indeed, the automated identification of such reports is not a new idea. There exist tools that classify issue reports as $non-issue$ or not, for reports written in English, such as TicketTagger \citep{kallis2019ticket, kallis2021predicting} or BEE \citep{song2020bee}. For the Tool Competition from the International Workshop on Natural Language-based Software Engineering (NLBSE'22) \citep{kallis2022nlbse}, teams submitted multiple classification models for the relevant task. However, these tools cannot be used to classify the reports for the industrial case, since the issue reports used in this study are written in Turkish, an agglutinative language (Section \ref{sec:morp}), where much of the semantic information is encoded by the suffixes or morphemes added to the words \cite{oflazer1994two, oflazer2014turkish}. A manual analysis of the issue reports revealed that there exist discourse patterns that signal $non-issue$ (Section \ref{sec:discourse_patterns}), which are rules to capture the syntax and semantics of the text \cite{chaparro2017detecting}. However, the automated extraction required morphological analysis, since most of the time the patterns are expressed by morphemes added to the verbs (Section \ref{sec:discourse_patterns}). Thus, before classification, we leverage morphological analysis to extract the roots and morphemes of the words. 

For the purpose of developing a tool to guide reporters during submission by automatically validating the report, as non-issue or not, we first randomly selected $1,200$ issue reports created in October 2020. We then labeled them as $non-issue$ or not, and extracted a list of patterns together with roots and suffixes to identify the discourse patterns (Table \ref{tab:patterns_nb}). Using feature extractors that we define as \textit{n-grams extractor}, \textit{morphological analysis-based extractor} and \textit{patterns-based extractor} (Section \ref{sec:classification}), we propose and evaluate a machine learning based method to predict the validity of incoming issue reports. 

The paper is organized as follows: Section \ref{sec:motivation} describes the industrial case and the motivation behind the study; Section \ref{sec:approach} presents the approach; Section \ref{sec:results} reports the results; Section \ref{sec:discussion} discusses the results; Section \ref{sec:related} presents the related work and Section \ref{sec:conclusion} concludes the paper. 

\section{Industrial Case and Motivation}
\label{sec:motivation}

Softtech received $111,534$ issue reports from IsBank in 2022, related to software problems, and typically concerning business critical systems. The reports, written in Turkish, include a one-line summary, a description, and often screenshot attachments. The issue reports are categorized into three levels based on their technical complexity. Level 1 reports are resolved by the IT Help Desk at IsBank, by using basic troubleshooting guides. If a resolution cannot be reached, the report is forwarded to Softtech. Level 2 reports are handled by the Application Support Team at Softtech, who resolve them if the issue does not require code base changes. If code analysis or modification is required, reports are assigned to software developers at Level 3. Of the total reports received in 2022, $87,214$ were solved at Level 2 and $24,320$ at Level 3.

At Softtech, issue reports assigned at Level 2 or 3 may be returned to the IT Help Desk at IsBank (Level 1) for various reasons such as, being a feature request or operational request, lacking information, or being outside of Softtech's scope of services. These reports are labeled as ``Returned''. Other labels in this category include ``Completed'', ``Cancelled'', or ``Unresolved''. In 2022, $28,775$ of the $111,534$ issue reports ($25.8\%$) were marked ``Returned'', while $79,496$ ($71.3\%$) were marked ``Completed''. 

We further compared the average resolution time and average bug tossing (or bug reassignment) count for these issue reports. It turns out that the average resolution time was $3.46$ days for ``Returned'' reports and $3.91$ days for ``Completed'' reports, meaning that ``Returned'' reports were nearly as costly as ``Completed'' reports in terms of time passing to solve them. Furthermore, the average bug tossing count was $0.74$ for ``Returned'' reports, and $0.34$ for ``Completed'' reports. So, the ``Returned'' reports were re-assigned more times than the ``Completed'' reports. 

Most of the issue reports ($23,072$) were returned at Level 2 by application support team members, while still some of them ($5,703$) could be detected at Level 3 by software developers, costing more. Furthermore, ``Returned'' reports were tossed more frequently at Level 3 than at Level 2. At Level 3, ``Returned'' reports were tossed $0.9$ times on the average, while at Level 2 they were tossed $0.7$ times. Additionally, average resolution times for ``Returned'' issue reports at Level 3 are longer than at Level 2 ($9$ and $2.09$, respectively), implying that the kinds of reports that the developers solve are harder and thus more costly. 

These results imply that detecting reports that should be returned to the reporter as early as possible is not an easy task, and such reports can get as costly as completing them by finding the problem and fixing it.

\section{Approach}
\label{sec:approach}

In this section, we describe our approach to develop a tool to automatically classify issue reports as $non-issue$ or not.

\subsection{Morphological Analysis}
\label{sec:morp}

Morphological analysis decomposes a word into its smallest possible functional units, its root and suffixes. It is an important concept for many NLP applications, such as spell checking, parsing, machine translation, dictionary tools and similar applications \cite{oflazer1994two}. Decomposing words into its smaller units is important especially for agglutinative languages such as Turkish or Finnish, where it is possible to express a whole sentence with only one word. Note that morphological analysis differs significantly from  stemming words. Stemming is a heuristic process that involves removing word endings in an attempt to identify the root, whereas morphological analysis properly performs this task by utilizing a vocabulary and returning the root in its dictionary form. 

Some one word examples in Turkish, with their possible roots and suffixes separated, are provided below. First the word, then the morphological breakdown, and then the English translation is provided. We used TRmorph\footnote{http://coltekin.net/cagri/trmorph/}, a tool to analyze words written in Turkish morphologically, to obtain the roots and suffixes. The words are selected from the relevant banking domain. 

\begin{enumerate}
  \item Bul-un-a-ma-dı / [Verb + Passive + Ability + Negative + Past Tense + Third Person Singular] / It could not be found.
  \item Çalış-tır-a-ma-dı-k / [Verb + Causative + Ability + Negative + Past Tense + First Person Plural] / We could not run it.
  \item Seç-meli-ydi / [Verb + Obligative + Past Tense + Third Person Singular] / It should have chosen.
\end{enumerate}

For all of the above examples, the root is a verb. Furthermore, they include patterns that are signs that the issue report including the word is a valid issue report. For all the examples, all the semantic information is included in the morphemes added to the verb. Without the morphological breakdown, it would not be possible to get the semantic information. 

For the first two examples, the negativity added to the verb with the suffix ``\textit{ma}'', indicates that the user is not able to perform a task, which indicates the \textit{observed behaviour} is expressed in the issue report, and the report can be classified as $issue$. For the third example, the suffix ``\textit{meli}'' added to the verb root ``\textit{seç (choose)}'' adds the meaning ``\textit{should}'', indicating the ``\textit{expected behaviour}'' is expressed in the issue report. As can be seen from the above examples, whole sentences can be expressed by simply adding suffixes to the roots in Turkish, and in order to classify the issue reports, these morphemes should be extracted first. In fact, in our studies we observed that most of the semantic information to check for, so as to classify as $non-issue$ or not, is encoded by the suffixes added to the verbs.

A well known two level morphological analyzer for Turkish was developed by Oflazer \cite{oflazer1994two}. In this work, Oflazer describes the rules and finite state machines for Turkish morphology. Based on this initial work, morphological analyzers were developed for Turkish such as TRmorph \cite{coltekin2010freely} or for Turkic languages such as Zemberek \cite{akin2007zemberek}. In our work, we used Zemberek \cite{akin2007zemberek} in automatic morphological analysis of the terms. 

\subsection{Data and Manual Labeling}

All the issues are reported through the Jira tool \footnote{https://jira.atlassian.com/} at Softtech and we use the Jira Api to extract the issue reports. We randomly selected $1,200$ issue reports created in October 2020, from the top ten products with the most number of issue reports in our domain, and manually analyzed them so as to classify as $non-issue$ or not, and to label the patterns signaling $non-issue$.  

For manual labeling, we initially selected $100$ issue reports out of $1,200$ and labeled them as $non-issue$ or not. If $non-issue$, we decided on the pattern that indicated $non-issue$. After deciding on the initial labels and related patterns, two authors of this paper labeled the remaining $1,100$ issue reports over a two-week period. The reports were presented to the labeling authors ensuring coverage across all ten products, with the help of a tool that we developed. The tool displays the issue report to be labeled, and the user selects whether to label it as $non-issue$ or not. If labeled as $non-issue$, the user then selects from the $non-issue$ patterns listed in the menu. To resolve disagreements between manual labelers, a daily meeting was held among the two manual labelers and the third author of this paper throughout the labeling process, to discuss the disagreements and come to a consensus on the labels. Also, if a new discourse pattern, indicating $non-issue$, is needed to be added to the catalog, they discussed it first, and added it to the database. This process continued until all of the $1,200$ issue reports were labeled. 

\subsection{Classification}
\label{sec:classification}

We consider the task of determining an issue report as $non-issue$ or not, as a supervised classification task. For this purpose, we employ a machine learning based approach. Specifically, we first score the terms in the corpus using \textit{tf-idf} scheme \citep{schutze2008introduction}, and then use LinearSVC \citep{joachims2005text} from scikit-learn \citep{pedregosa2011scikit} for classification, which performed best compared to other classification techniques in our previous studies \citep{aktas2020automated, aktas2022using} on automated issue report assignment at Softtech. 

We use three feature extractors to employ this pipeline. The first one, which we call \textit{n-grams extractor} (or shortly \textit{n-grams}), simply tokenizes the words in the issue reports, removes the special characters and stop-words, and returns the extracted n-grams, where $n={1, 2}$ in our case. The second extractor is called \textit{morphological analysis-based extractor} (or shortly \textit{ma}). It first determines the verbs in the text, then morphologically breaks down the verb into its morphemes, finally returns these morphemes. The third extractor is called \textit{pattern-based extractor}, shortly \textit{patterns}. It basically returns the matching parts of the morphologically analyzed verb, after the \textit{ma} extractor returns the morphemes. Note that \textit{ma} and \textit{patterns} uses the semantic representations of the suffixes rather than their lexical representations, since the same semantic representation may differ lexically in Turkish. We utilize the Zemberek tool \citep{akin2007zemberek} to identify the verbs, and then extract their roots and the types of suffixes added to these roots. 



We use the $1,200$ issue reports in our experiments and use cross validation, where we divide the dataset into 10 folds and reserve one fold for testing. We repeat this process following the 10-fold cross-validation approach, and report the average values of the evaluation metrics. 

We evaluate the models using $precision$, $recall$ and $F1-score$ \citep{dougherty2012pattern}. Precision is the proportion of correct predictions out of the total instances predicted as positive, while recall is the proportion correct predictions out of the total actual positive instances. F1-score is a harmonic mean of precision and recall, thus providing a balanced assessment of the model's performance. As a result of the evaluation, we select and build a binary classifier for the classification task as $non-issue$ or not. 

\section{Results}
\label{sec:results}

This section presents the results of manual labeling and the experiments related to $non-issue$ classification.

\subsection{Distribution of Non-Issue Reports}

Out of the $1,200$ manually labeled issue reports, $159$ were classified as $non-issue$. The selected ten projects and the distribution of $non-issue$ reports among them are given in Table \ref{tab:data}. The project with the highest percentage of $non-issue$ was Credit Cards ($25.48\%$), while Consumer Loan Services had no $non-issue$. 

\begin{table}
  \caption{The distribution of non-issue reports among projects (with related percentages in parentheses).}
  \label{tab:data}
  \begin{tabular}{|l|cc|c|}
    \hline
    \textbf{Project} & \textbf{\#Non-Issue} & \textbf{\#Issue} & \textbf{Total}\\
    \hline
    Bancassurance & 18 (15.79\%) & 96 & \textbf{114} \\
    \hline
    Collateral Management & 13 (11.93\%) & 96 & \textbf{109} \\
    \hline
    Commercial Loans& & & \\
    \hline
    Allocation & 9 (5.06\%) & 169 & \textbf{178} \\
    \hline
    Commercial Loans & & & \\
    Disbursement & 1 (1.02\%) & 97 & \textbf{98} \\
    \hline
    Consumer Loan Services & 0 (0\%) & 76 & \textbf{76} \\
    \hline
    Credit Cards & 40 (25.48\%) & 117 & \textbf{157} \\
    \hline
    Customer Information & & & \\
    Management & 22 (21.78\%) & 79 & \textbf{101} \\
    \hline
    Deposits & 5 (7.04\%) & 66 & \textbf{71} \\
    \hline
    Foreign Currency Transfers & 32 (21.19\%) & 119 & \textbf{151} \\
    \hline
    Personal Loans Allocation & 19 (13.10\%) & 126 & \textbf{145} \\
    \hline
    \textbf{Total} & \textbf{159 (13.25\%)} & \textbf{1041} & \textbf{1,200} \\
    \hline
  \end{tabular}
\end{table}

\begin{table*}
  \caption{Non-issue discourse patterns, with English translations given in parenthesis.}
  \label{tab:patterns_nb}
  \begin{tabular}{|l|ll|l|}
    \hline
    \multirow{2}{*}{\textbf{Pattern Code}}&
    \multicolumn{2}{|c|}{\textbf{Rule}} & \multirow{2}{*}{\textbf{Example sentence/part of sentence}} \\
    \cline{2-3}
    & \textbf{Word/Root} & \textbf{Suffix} & \\
    \hline
    NI\_REQUEST & arşiv, güncel, sil & -mesi/-ması & Başvurunun \textbf{arşiv}e kaldırıl\textbf{ması}nı rica ederiz. \\
    & (archive, update, delete) & - & (We request the application to be archived.) \\    
    \hline
    NI\_YESNO\_QUESTION & - & -mi/-mı & ... poliçe numarasını öğrenebilir \textbf{mi}yiz? \\
    & - & - & (Can we learn the policy number ...?) \\
    \hline
    NI\_WHY\_QUESTION & neden (why) & - & \textbf{Neden} hayat sigortası polisinin iptal edildiği ... \\
    & - & - & (Why the life insurance policy was canceled ...?) \\
    \hline
    NI\_POSSIBLE & mümkün (possible) & - & ... \textbf{mümkün} mü? \\
    & & & (Is it possible to ...?) \\ 
    \hline
    NI\_INADVERTENTLY & sehven (inadvertently) & - & ... nolu aday müşterinin silinmesi talep \\
    & & & edilmektedir. \textbf{Sehven} yaratılmıştır. \\
    & & & (Deletion of the candidate customer number \\ 
    & & & ... is requested. It was created inadvertently.) \\
    \hline
  \end{tabular}
\end{table*}

The distribution of $non-issue$ reports is not uniform across projects, with one project having around a quarter of its reports labeled as $non-issue$, while another project has none. This disparity may be due to differences in how software teams use the issue reporting channel to communicate with customers. Some teams may not consider $non-issue$ reports to be within the scope of this channel, while others may use it more broadly to communicate various issues. Additionally, for some teams, the complexity of their systems (e.g. Credit Cards infrastructure) may lead them to implement workarounds for certain tasks instead of developing permanent solutions for incoming $non-issue$ reports. 

In conclusion, $13.25\%$ of the issue reports are labeled as $non-issue$, and their distribution varies from one project to another. This indicates that there may be a lack of clarity or understanding among users regarding the issue reporting process. This finding, together with the costs regarding the ``Returned'' issue reports we presented in Section \ref{sec:motivation}, could prompt Softtech to develop a tool to warn the reporters before the submission of an issue. 

\subsection{Non-Issue Discourse Patterns}
\label{sec:discourse_patterns}

The list of $non-issue$ discourse patterns extracted with manual analysis are given in Table \ref{tab:patterns_nb}. The most common $non-issue$ pattern is NI\_REQUEST, where the user requests an operational change or correction. This pattern is observed in $93.29\%$ of the $non-issue$ reports. Some example sentences are: \textit{``We require to know if there were any address updates between the dates ...'', ``We request the following credit application to be archived.''.} Such requests indicate that the users may not have the necessary screens to perform the desired operation, or they do not know how to perform it. Other common types of $non-issue$ patterns are NI\_YESNO\_QUESTION and NI\_WHY\_QUESTION ($6.71\%$), where the user asks for a reason or information. Examples are: \textit{``We have corrected the card status as ... Why was it ... before?''.}, \textit{``Can we learn the policy number that was collected and returned on ... from the credit card ... ?}. Also some examples of NI\_INADVERTENTLY is observed, where the user tells that some operation is performed by mistake and asks the software team to fix it.  

To automate the detection of NI\_REQUEST, the verb roots such as ``archive'', ``update'' and ``delete'' are required to be extracted. They are frequently used with the suffix ``-mesi'', ``-ması''. The suffixes ``-mı, -mi'' added at the end of the sentences are also needed to be detected, since they are used to ask a yes/no question in Turkish. 

The results confirm the existence of specific discourse patterns for $non-issue$ detection, supporting the development of an automated tool to identify them. The patterns require morphological analysis for identification, as they are identified by verb roots and/or suffixes. Note also that, as we gave some examples in Section \ref{sec:morp}, there exist patterns in valid issue reports as well, such as negative verbs having suffixes ``-me, -ma'' which express the observed behaviour of the software; or obligating suffixes such as ``-meli, -malı'' expressing the expected behaviour.

\subsection{Detecting Issue Report Validity}

\begin{table}
  \caption{Average (P)recision, (R)ecall and (F)-measure for detection of non-issue in manually labeled issue reports.}
  \label{tab:classification}
  \begin{tabular}{|l|ccc|}
    \hline
    \textbf{Features} & \textbf{P} & \textbf{R} & \textbf{F} \\
    \hline
    \textit{n-grams} & 0.70 & 0.54 & 0.59\\
    \textit{ma} & 0.67 & \textbf{0.78} & 0.71\\
    \textit{patterns} & 0.33 & 0.60 & 0.42\\
    \textit{n-grams} + \textit{ma} & \textbf{0.80} & 0.73 & 0.76\\
    \textit{n-grams} + \textit{patterns} & 0.74 & 0.56 & 0.62\\
    \textit{ma} + \textit{patterns} & 0.64 & 0.73 & 0.73\\
    \textit{n-grams} + \textit{ma} + \textit{patterns} & 0.78 & 0.76 & \textbf{0.77}\\
  \hline
  \end{tabular}
\end{table}	

We have defined three feature extractors in Section \ref{sec:classification} for the classification of issue reports: \textit{n-grams}, \textit{ma} and \textit{patterns}. We have used all the combinations of these extractors in our experiments, where our aim was to see the individual effects the extractors. After extracting the features, we represent them with \textit{tf-idf}, and classify using LinearSVC algorithm. The n-gram range is $(1, 2)$ for the tf-idf vectorizer, and the regularization parameter $C$ of the LinearSVC algorithm is set to 1. Table \ref{tab:classification} presents the results. 

The results show that any approach that utilizes morphological analysis \textit{ma} achieves an F1-score of at least $0.71$. On the other hand, the use of only \textit{patterns} leads to an F1-score of $0.42$ suggesting that defining the patterns in valid issue reports and using them in classification can improve the performance with the \textit{patterns} extractor. 

The best F1-score ($0.77$) is achieved when we combine the features extracted by \textit{patterns}, \textit{ma} and \textit{n-grams}. We select this method as the proposed approach in this paper. In conclusion, incorporating morphological analysis to extract roots and suffixes as input features improved the classification results.

\section{Discussion}
\label{sec:discussion}

Validating issue reports in an industrial setting is different than validating them for open source projects (such as the ones in GitHub). In GitHub, an issue is normally resolved in longer time (weeks, or even months), while in an issue-triaging scenario in an industrial setting, the time to resolve is typically a few days. Also, issue reports in an industrial setting differ from open source projects, due to the heavy use of special terms related to the relevant industry, and the language valid in the country where the company is located. As a result, the discovered discourse patterns may be specific for the industrial case. 

As the best F1-score in our case is currently $0.77$, the stakeholders at Softtech and IsBank may have concerns regarding the accuracy of the proposed system. When deploying the model at Softtech, a gradual transition strategy could be followed to address the rightful concerns of the stakeholders. If an issue report is detected as $non-issue$ by the machine learning model, the Help Desk could be asked to verify that the report is to be returned immediately to the reporter. As the stakeholders build confidence in the system, a confidence factor could be attached to each prediction so that $non-issue$ reports with high confidence are returned without asking the Help Desk to verify. 

\section{Related Work}
\label{sec:related}

Our findings at Softtech align with previous research such that a significant portion of the issue reports are non-issues \cite{antoniol2008bug, pingclasai2013classifying, herzig2013s, zhou2016combining, pandey2017automated, terdchanakul2017bug, qin2018classifying, kallis2019ticket, otoom2019automated, herbold2020feasibility, he2020deep, xie2021mula, perez2021bug, izadi2022predicting}. 

For instance, Antoniol et al. \cite{antoniol2008bug} conducted a study aimed at developing an automated classifier to distinguish between corrective maintenance issues, and the other types of issues. To build the classifier, they collected $1,800$ issue reports from open source projects, Mozilla, Eclipse, and JBoss. These reports were manually classified as bug or non-bug. According to the manual classification, $45\%$ of issue reports in Mozilla, $32\%$ in Eclipse and $\%58$ in JBoss were categorized as bugs, while the remaining reports were classified as non-bug. Using the relevant words for bug/non-bug classification and tf-idf vectorization, the researchers achieved accuracies between $0.77$ and $0.82$ using Naive Bayes, ADTree, and Logistic Regression classifiers.

To reduce the valuable developer time spent on labeling issue reports as bug or non-bug, Pandey et al. \cite{pandey2017automated} developed a tool that automatically classifies the reports. They employed a manually classified corpus of $5,587$ issue reports from three open-source projects: HttpClient, Lucene, and Jackrabbit of Apache. Of these, $305$ out of $745$ reports from HttpClient, $697$ out of $2,441$ reports from Lucene, and $937$ out of $2,401$ reports from Jackrabbit were classified as bugs, while the remainder were classified as non-bug. To classify the reports, the researchers converted each text into a document term matrix using the summary as input. They tested various algorithms for classification, including Naive Bayes, Linear Discriminant Analysis, K-Nearest Neighbors, Support Vector Machine (SVM) with various kernels, Decision Tree, and Random Forest. Depending on the project, Random Forest and SVM achieved the best results with an accuracy ranging from $0.75$ to $0.83$.

Ticket Tagger is a GitHub app that automatically labels issue reports in any GitHub repository's issue tracker as bug, enhancement, or question, as introduced by Kallis et al. \cite{kallis2019ticket, kallis2021predicting}. The researchers utilized the subject and description features of the reports as input, represented them as bag of n-grams, which is a set of sequences of n consecutive words, and classified them using \textit{fastText} \citep{joulin2016bag}, a multi-class linear neural model. To evaluate the tool's performance, they tested it on $30,000$ issue reports sampled from $12,112$ GitHub projects, distributed evenly among the three categories. The 10-fold cross-validation F1-score results indicated an accuracy of $0.83$, $0.82$, and $0.83$ for bugs, enhancements, and questions, respectively.

In their study, Herbold et al. \cite{herbold2020feasibility} investigated the feasibility of using machine learning techniques to automatically classify issue reports as bugs or non-bugs. To conduct their experiments, they employed four datasets comprising over $600,000$ issue reports and replicated several existing machine learning approaches from the literature. By comparing their performance with more data, they found that the models achieved high F1-scores, demonstrating the potential of issue type prediction tools for practical use.

In bug report validation, deep learning-based techniques have been gaining attention lately. In their study, He et al. \cite{he2020deep} proposed a CNN-based architecture that utilizes only textual features to predict the validity of bug reports collected from five open-source projects, including Eclipse, Netbeans, Mozilla, Thunderbird, and Firefox. They also identified various bug report patterns that indicate validity, which vary across projects. On average, they achieved F1-scores of $0.80$ and $0.69$ for valid and invalid bug reports, respectively, across the five projects.

In their study, Trautsch et al. \cite{trautsch2022predicting} employed a pre-trained language model called seBERT, which is based on the BERT (Bidirectional Encoder Representations from Transformers) architecture \cite{devlin2018bert} and trained using software engineering data, to predict issue types, namely bug, enhancement, or question. They utilized issue reports from GitHub, as used in previous studies \citep{kallis2019ticket, kallis2021predicting, kallis2022nlbse}, with $722,899$ reports for training and $80,518$ for testing. They achieved an overall F1-score of $0.86$, which outperformed the baseline fastText model.

The International Workshop on Natural Language-based Software Engineering (NLBSE’22) organized a tool competition to classify issue reports as bugs, enhancements or questions \cite{kallis2022nlbse}. The dataset included more than $803,417$ issue reports from $127,595$ open source projects from GitHub. Five teams competed with multiple classification models \cite{izadi2022catiss, bharadwaj2022github, colavito2022issue, siddiq2022bert, trautsch2022predicting}, where most of them are based on the BERT architecture \cite{devlin2018bert}. All the classifiers achieved high performances as a result, however, CatIss method by Izadi \cite{izadi2022catiss} performed the best. Additional pre-preprocessing steps of the issue reports appeared to be effecting the most for such high performances. 

Compared to previous work, we conduct our study in an industrial set-up where the issue reports are written in Turkish and in banking domain. We hypothesize that discourse patterns also exist in this domain, and manually label the issue reports, and analyze them to discover these patterns. Since Turkish is an agglutinative language, we integrate morphological analysis in our work as a feature extractor. As a result, we show that utilizing these patterns, and relevant morphemes as input features enhances the model's performance compared to using only n-grams. 

\section{Conclusion}
\label{sec:conclusion}

In this work, we aimed to develop a tool in an industrial setting, that guides the reporters or the Help Desk to decide on the validity  of incoming issue reports, since returning the reports by the application support team members or the software developers is more costly. 

For this purpose, we selected and manually analyzed $1,200$ issue reports from the banking domain. We labeled them as $non-issue$ or not, and manually extracted the patterns indicating $non-issue$. As a result, $13.25\%$ of the reports are labeled as $non-issue$ and $5$ discourse patterns are defined. The proposed approach performed a $0.77$ F1-score at best, while morphological analysis improved validity detection in all settings. As future work, we plan to extend our work by using other prediction models such as the ones that use BERT architecture \cite{izadi2022catiss, bharadwaj2022github, colavito2022issue, siddiq2022bert, trautsch2022predicting}. 

As validating issue reports in an industrial setting is different than validating them for open source projects, the results could be specific for the industrial case. Also, a gradual transition strategy could help the stakeholders build confidence in the automated validation system. 

In conclusion, the results indicate that validating issue reports can get costly as the level of technical complexity to resolve the problem increases, the distribution of $non-issue$ reports may vary among the software teams depending on the way they use the triaging process to communicate with the customers, there exist discourse patterns that allow us to understand if the reports are valid or not, and an automated method to validate the issue reports will improve the process for the industrial case. 

\bibliographystyle{ACM-Reference-Format}
\bibliography{sample-base}

\end{document}